%% file: Kucera_V.tex
\documentclass[3p,times,twocolumn]{elsarticle}
\pdfoutput=1

\usepackage{ecrc}

\input{Kucera_V_header.tex}

\volume{00}

\firstpage{1}

\journalname{Nuclear Physics B Proceedings Supplement}

\runauth{V. Kučera (ALICE Collaboration)}

\jid{nppp}

\jnltitlelogo{Nuclear Physics B Proceedings Supplement}

\usepackage{lineno}

\usepackage[figuresright]{rotating}

\input{Kucera_V_commands.tex}

\begin{document}

\begin{frontmatter}



\dochead{}

\title{Production of strange particles in charged jets in \pbpb{} and \ppb{} collisions measured with ALICE}


\author[1,2]{Vít Kučera (for the ALICE Collaboration)}

\address[1]{Nuclear Physics Institute of the CAS, v. v. i., Na Truhlářce 39/64, 180 86 PRAHA 8, Czech Republic}
\address[2]{IPHC, 23 rue du Loess - BP 28, F-67037 STRASBOURG CEDEX 2, France}

\ead{vit.kucera@cern.ch}

\begin{abstract}
\input{Kucera_V_abstract.tex}
\end{abstract}

\begin{keyword}


ALICE, heavy ions \sep jets \sep fragmentation \sep strange particles \sep baryon-to-meson ratio \sep baryon anomaly

\end{keyword}

\end{frontmatter}

%\linenumbers

\input{Kucera_V_proceedings.tex}



%% References

%\nocite{*}
\bibliographystyle{elsarticle-num}
\bibliography{Kucera_V_bibliography}






% \end{thebibliography}

\end{document}

%% file: Kucera_V_header.tex
\usepackage[utf8]{inputenc} % source file encoding
\usepackage[czech,greek,british]{babel} % used languages (last one is default)
\usepackage{amsmath,amsthm,amssymb} % AMS advanced mathematics (\text), AMS theorems, AMS symbols+fonts (\lesssim)
\usepackage{isomath} % correct shapes of math symbols according to ISO standards (greek symbols, bold vectors, tensors, etc.)
\usepackage{siunitx} % SI units, correct typesetting of quantities

%% file: Kucera_V_commands.tex
\newcommand{\gr}[1]{{\foreignlanguage{greek}{#1}}} % default style of the greek font
\newcommand{\urlss}[1]{\hbox{\scriptsize \url{#1}}} % URLs with small font size

\newcommand{\pic}{\textnormal{\gr{p}}} % constant pi
\newcommand{\particle}[1]{\textnormal{#1}} % all particles must be typeset using an upright non-bold font
\newcommand{\partgr}[1]{{\particle{\gr{#1}}}} % particle denoted by a greek letter, not available at 6 pt in math mode
\newcommand{\pt}{{p_\textnormal{T}}} % transverse momentum (pt)
\newcommand{\ptjch}{{p_\textnormal{T}^\textnormal{jet,ch}}} % pt of a charged jet
\newcommand{\etavo}{{\eta_{\vo}}} % eta of a V0 particle
\newcommand{\ajch}{{A_\textnormal{jet,ch}}} % area of a charged jet
\DeclareSIUnit\clight{\ensuremath{c}}
\newcommand{\mc}{\si[per-mode=symbol]{\mega\electronvolt\per\clight}} % MeV/c
\newcommand{\gc}{\si[per-mode=symbol]{\giga\electronvolt\per\clight}} % GeV/c
\newcommand{\tev}{\si{\tera\electronvolt}} % TeV
\newcommand{\kt}{\ensuremath{k_\textnormal{t}}} % k_t algorithm, warning: {} needed after the cmd if another text follows after
\newcommand{\akt}{anti-\kt} % anti-k_t algorithm
\newcommand{\snn}{\ensuremath{\sqrt{s_\particle{NN}}}} % cms energy per nucleon pair
\newcommand{\pbpb}{\mbox{\pb--\pb}} % symbol for lead-lead, correct and ALICE convention
\newcommand{\ppb}{\mbox{\proton--\pb}} % symbol for lead-lead, correct and ALICE convention
\newcommand{\pp}{\proton\proton} % symbol for proton-proton, wrong but ALICE convention
\newcommand{\interact}{\ensuremath{\rightarrow}} % symbol for interaction equations

\newcommand{\anti}[1]{{\ensuremath{\overline{#1}}}} % anti-particle

\newcommand{\pb}{\particle{Pb}} % lead (element or nucleus)
\newcommand{\vo}{{\ensuremath{\particle{V}^\textnormal{0}}}}
\newcommand{\proton}{{\particle{p}}}
\newcommand{\pion}{\partgr{p}}
\newcommand{\kaon}{\particle{K}}

\newcommand{\kos}{\ensuremath{{\kaon^\textnormal{0}_\textnormal{S}}}}
\newcommand{\lmb}{\partgr{L}}

%% file: Kucera_V_abstract.tex
Measurements of spectra of identified particles produced in jets represent an important tool for understanding the interplay of various hadronisation mechanisms which contribute to particle production in the hot and dense medium created in ultra-relativistic heavy-ion collisions.
In this contribution, we present the measurements of the $\pt$~spectra of \lmb{}~baryons and \kos{}~mesons produced in charged jets in \pbpb{} collisions at $\snn=\SI{2.76}{\tev}$ and in \ppb{} collisions at $\snn=\SI{5.02}{\tev}$.
The results are obtained with the ALICE detector at the LHC, exploiting the excellent particle identification capabilities of this experiment.
Baryon-to-meson ratios of the spectra of strange particles associated with jets are studied in central collisions and are compared with the inclusive ratios.

%% file: Kucera_V_proceedings.tex
\section{Introduction}

An enhancement of the baryon-to-meson ratio has been  observed for inclusive production of light-flavour particles at intermediate transverse momenta ($\SI{2}{\gc} < \pt < \SI{6}{\gc}$) in \mbox{heavy-ion} collisions when compared to the ratio measured in proton--proton (\pp) collisions.
The effect was first observed at the Relativistic Heavy Ion Collider~\cite{star-ratio-p-pi-auau,star-ratio-L-K-auau,star-ratio-L-K-auau-2}
 and later at the Large Hadron Collider (LHC) as well~\cite{alice-spectra-LK-pbpb}.
The ratio of the inclusive $\pt$ spectrum of \lmb{}~baryons to the spectrum of \kos{}~mesons measured in lead--lead (\pbpb{}) collisions with the ALICE experiment increases strongly with centrality and, for the most central collisions, reaches a~maximum three times higher than the ratio obtained for \pp{} collisions.
A smaller but still significant enhancement is manifest in proton--lead (\ppb) collisions as well~\cite{alice-spectra-light-ppb}.
This phenomenon is not understood yet and various mechanisms have been proposed to explain it.

One of considered explanations is modification of jet fragmentation in medium.
Jet fragmentation denotes the process of hadron production from a~parton with large momentum produced in a~hard scattering.
The energetic parton undergoes parton showering followed by hadronisation which gives origin to a~collimated spray of hadrons called a~jet.
The process might be sensitive to interactions of fragmenting partons with the strongly-interacting matter created in ultra-relativistic heavy-ion collisions.
Such interaction might modify properties of the resulting jet, including $\pt$ spectra of the hadrons which constitute the jet.

Some attempts to explain the enhancement of the baryon-to-meson ratio propose contribution from various scenarios based on hadronisation by parton recombination which assumes that in a~dense medium, where phase space is filled with partons, hadrons are produced by the clustering of multiple partons together~\cite{hydro-and-hybrid-cascade,coalescence-models,epos,hwa-recombination-minijet-2,hwa-recombination-minijet-3}.
That would result in a~harder $\pt$ spectrum of baryons with respect to a~spectrum of mesons.

The goal of the presented analysis is to study the origin of the enhancement by separating hadrons produced in association with hard processes and hadrons produced in the bulk.
The results of this analysis aim to disentangle contributions of processes in bulk and potential contribution of modified jet fragmentation in medium.

\section{Analysis}

%\subsection{Setup}

The analysis is performed on data recorded with the ALICE detector at the LHC during the runs with \pbpb{} and \ppb{} collisions at the centre-of-mass energies of $\snn=\SI{2.76}{\tev}$ and $\SI{5.02}{\tev}$, respectively.
Tracking of charged particles in the central barrel is provided by the Inner Tracking System and the Time Projection Chamber, both placed in a~magnetic field of $\SI{0.5}{\tesla}$.
The centrality of collisions is estimated from the multiplicity of charged particles measured in the V0~detectors at forward pseudorapidities~\cite{alice-ppr-1}.

The analysis of neutral strange particles in charged jets consists of two main parts: reconstruction of charged jets and reconstruction of neutral strange particles.

%\subsection{Analysis of charged jets}

The jet reconstruction is performed using primary charged particles with $\pt$ greater than \SI{150}{\mc} within the pseudorapidity acceptance of \mbox{$|\eta_\textnormal{track}| < 0.9$}.
Charged particles are clustered into jets by the \akt{} jet algorithm (implemented in the FastJet package~\cite{fastjet}) using resolution parameter \mbox{$R = 0.2$} for \pbpb{} collisions and $R = 0.2, 0.3, 0.4$ for \ppb{} collisions.

The jet algorithm cannot distinguish jet fragments from particles produced in the underlying event.
The mean density of background contributing to the reconstructed jet momenta is estimated in each event from clusters reconstructed with the \kt{} algorithm~\cite{alice-jets-fluctuations-pbpb}.
The corresponding momentum is then subtracted for each reconstructed jet.
In order to increase the probability of selecting a hard scattering in \pbpb{} collisions, thresholds are imposed on jet $\pt$ ($\ptjch$), $\pt$ of the leading constituent of the jet (\mbox{$p_\textnormal{T}^\textnormal{leading track}>\SI{5}{\gc}$}) and jet area~(\mbox{$\ajch>0.6\pic R^2$}).
Spectra of strange particles in charged jets are studied for two thresholds of jet momentum: \mbox{$\ptjch>\SI{10}{\gc}$} and \mbox{$\ptjch>\SI{20}{\gc}$}.

%\subsection{Analysis of neutral strange particles}

Neutral strange particles (\vo) are reconstructed using the topology of their most probable weak decays into charged particles, namely \mbox{$\kos\interact\pion^++\pion^-$}, \mbox{$\lmb\interact\proton+\pion^-$} and \mbox{$\anti{\lmb}\interact\anti{\proton}+\pion^+$}.
Combinatorial background from fake \vo{}~candidates is strongly suppressed by cuts applied on parameters of the decay and the signal yields are extracted from the resulting invariant-mass distribution.
Spectra of \anti{\lmb}~baryons are analysed separately from those of \lmb{}~baryons and combined into the baryon-to-meson ratio as $(\lmb+\anti{\lmb})/2\kos$.

%\subsection{Strange particles in jets}

If a~\vo{}~candidate fulfils the selection criteria, the angular distance between its momentum vector and the axis of each selected jet in the event is calculated.
A~jet is considered for matching with \vo{}~candidates only if its cone is fully contained within the acceptance of \vo{}~particles: $|\eta_\textnormal{jet,ch}| < |\etavo|^\textup{max} - R$.
The \vo{}~candidate is considered to be inside the jet cone if its distance to the jet axis is less than the resolution parameter of the jet finder:
\begin{linenomath}
\begin{equation}
\sqrt{(\varphi_{\vo}-\varphi_\textnormal{jet,ch})^{2}+(\etavo-\eta_\textnormal{jet,ch})^{2}} < R.
\end{equation}
\end{linenomath}

%\subsection{Corrections}

\vo{}~particles collected within jet cones originate not only from jet fragmentation but also from the underlying event.
Spectra of particles coming from the underlying event are estimated by several methods where \vo{}~candidates are associated with regions where only contribution from the underlying event is expected.

An efficiency correction is applied to raw \vo{}~spectra using the inclusive simulated particles.
Then, spectra of particles in the underlying event are subtracted from spectra of particles in jet cones.
The resulting spectra still contain a~contamination from decays of jet constituents.
The corresponding fraction is subtracted by applying a~correction for feed-down from decays of $\partgr{X}^{0}$~and $\partgr{X}^{-}$~baryons ($\partgr{X}^{0,-}$) into \lmb{}~baryons.
Since the spectra of $\partgr{X}^{0,-}$~baryons in jets have not been measured yet, the feed-down fraction of inclusive \lmb{}~baryons is used as the default option.
To estimate the related systematic uncertainty, we use spectra of hyperons in jets in simulated \pp{} collisions at \mbox{$\sqrt{s}=\SI{2.76}{\tev}$} or $\SI{5.02}{\tev}$ generated by \mbox{PYTHIA~8}, tune 4C~\cite{pythia6,pythia8}.

\section{Results}

\begin{figure}[htbp]
\centering
\includegraphics[width=0.5\textwidth]{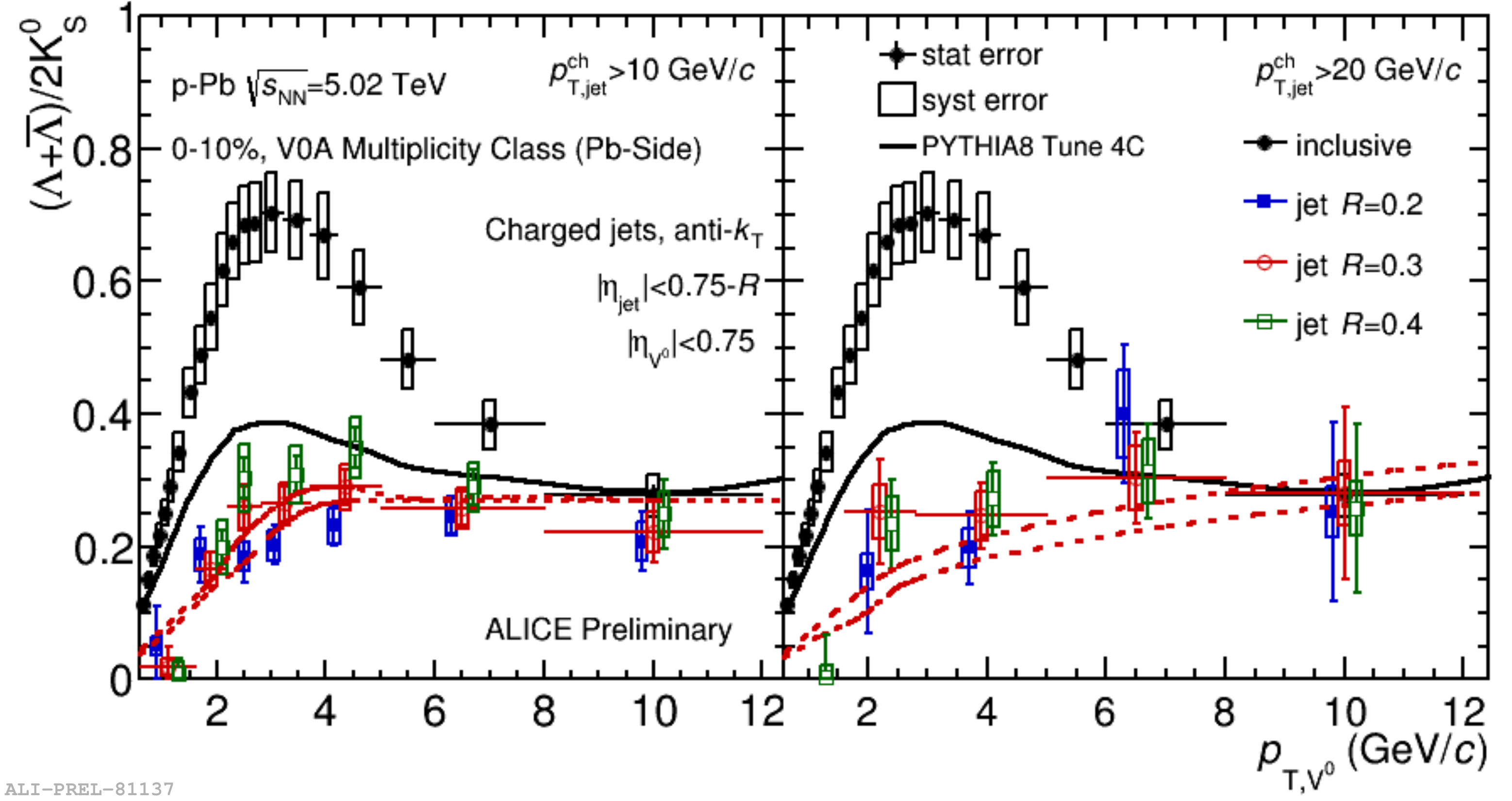}
\caption{$\lmb/\kos$ ratio in charged jets with $R=0.2,0.3,0.4$ in \ppb{} collisions at \mbox{$\snn=\SI{5.02}{\tev}$} for \mbox{$\ptjch>\SI{10}{\gc}$} (left) and \mbox{$\ptjch>\SI{20}{\gc}$} (right), compared with the inclusive ratio in \ppb{} and simulation results in \pp. Black solid line indicates the inclusive ratio from \mbox{PYTHIA~8}. Red dashed lines denote the spread of ratios in PYTHIA jets for all used values of~$R$.}
\label{fig:ratio-jets-ppb}
\end{figure}

Figure~\ref{fig:ratio-jets-ppb} shows the ratio of the $\pt$ spectra of \lmb{}~and \anti{\lmb}~baryons to that of \kos{}~mesons measured in charged jets in high-multiplicity \ppb{} collisions for two $\ptjch$ thresholds.
The ratio is compared with the measured inclusive ratio and with ratios obtained from simulations performed with \mbox{PYTHIA~8}.
The measured baryon-to-meson ratio in jets in \ppb{} collisions is below the inclusive ratio i)~measured in \ppb{} collisions, ii)~measured in \pp{} collisions~\cite{alice-spectra-LK-pp} (not shown) and iii)~obtained with PYTHIA simulations.
Moreover, the measured ratio exhibits a~surprising similarity with ratios of particles in jets simulated in PYTHIA and does not evince any significant dependence on $R$~or $\ptjch$.
The results indicate no visible modification of strangeness production in charged jets in \ppb{} compared to \pp{} and the enhancement of the baryon-to-meson ratio thus seems to be due to effects of the underlying event.

\begin{figure}[htbp]
\centering
\includegraphics[width=0.5\textwidth]{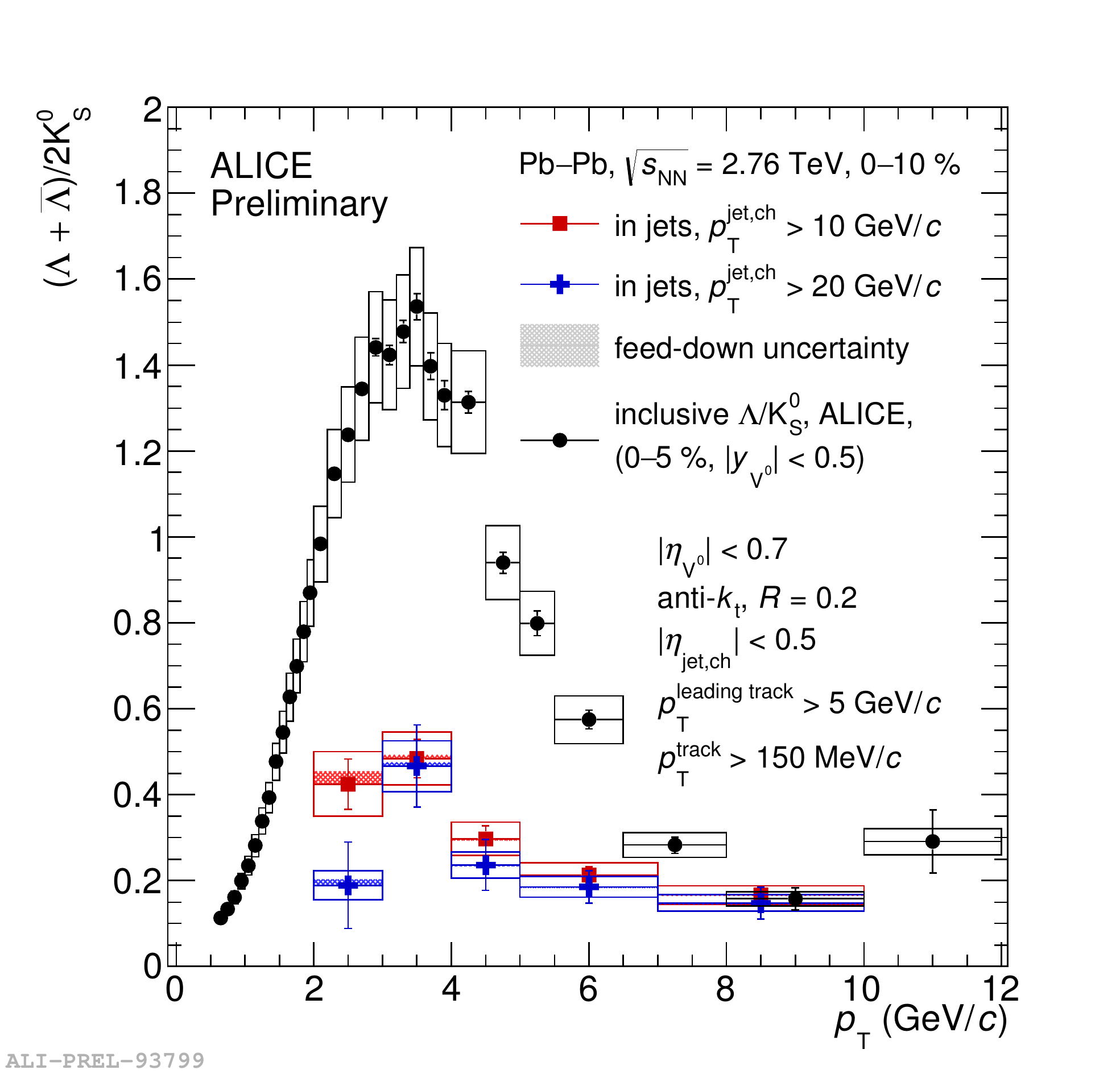}
\caption{$\lmb/\kos$ ratio in charged jets with $R=0.2$ in \pbpb{} collisions at \mbox{$\snn=\SI{2.76}{\tev}$} for \mbox{$\ptjch>\SI{10}{\gc}$} and \mbox{$\ptjch>\SI{20}{\gc}$}, compared with the inclusive ratio.}
\label{fig:ratio-jets-pbpb}
\end{figure}

Figure~\ref{fig:ratio-jets-pbpb} shows the $\pt$ dependence of the $\lmb/\kos$ ratio measured in charged jets in central \pbpb{} collisions, compared with the inclusive ratio measured in centrality range $\SIrange{0}{5}{\percent}$.
The ratio measured for particles in jets is significantly lower than the inclusive ratio at intermediate $\pt$ without exhibiting any dependence on the $\ptjch$ threshold.
The ratio in jets and the inclusive ratio meet at higher $\pt$ where production by jet fragmentation starts to be the dominant hadronisation process.

\section{Summary}

We described techniques used in the analysis of the $\lmb/\kos$ ratio in charged jets.
We presented the first measurement of the spectra of \lmb{}~baryons and \kos{}~mesons in jets in \pbpb{} and \ppb{} collisions at the LHC, performed by ALICE.

In both collision systems, the production of strange particles associated with jet fragmentation differs significantly from inclusive production and does not seem to depend on the minimum jet momentum.
The measured $\lmb/\kos$ ratios indicate that the enhancement does not originate from modified jet fragmentation for the given sample of jets.
%originates from bulk effects.

\section*{Acknowledgement}

The work has been supported by the grants LG13031 and 7AMB14FR066 of the Ministry of Education of the Czech Republic.